# Classification of multi-frequency RF signals by extreme learning, using magnetic tunnel junctions as neurons and synapses

Nathan Leroux[1], Danijela Marković[1], Dédalo Sanz-Hernández[1], Juan Trastoy[1], Paolo Bortolotti[1], Alejandro Schulman[2], Luana Benetti[2], Alex Jenkins[2], Ricardo Ferreira[2], Julie Grollier[1] and Alice Mizrahi[1*]

[1]Unité Mixte de Physique CNRS, Thales, Université Paris-Saclay, 91767 Palaiseau, France
[2]International Iberian Nanotechnology Laboratory (INL), 4715-31 Braga, Portugal
*alice.mizrahi@cnrs-thales.fr

Extracting information from radiofrequency (RF) signals using artificial neural networks at low energy cost is a critical need for a wide range of applications from radars to health. These RF inputs are composed of multiples frequencies. Here we show that magnetic tunnel junctions can process analogue RF inputs with multiple frequencies in parallel and perform synaptic operations. Using a backpropagation-free method called extreme learning, we classify noisy images encoded by RF signals, using experimental data from magnetic tunnel junctions functioning as both synapses and neurons. We achieve the same accuracy as an equivalent software neural network. These results are a key step for embedded radiofrequency artificial intelligence.

**Introduction**

Analyzing radiofrequency (RF) signals is essential in various applications, such as connected objects, radar technology, gesture sensing and biomedical devices[1–8]. For many signal classification tasks, for instance emitter type identification, artificial neural networks have proven to perform better than standard methods and show superior robustness to noise and defects[1]. However, running neural networks on conventional computing hardware can be time-consuming and energy-intensive, which makes it challenging to integrate this capability into embedded systems[9,10]. This issue is amplified in the case of RF signals, because they require signal digitization before being processed by the neural network.

A promising path to reduce the energy consumption of artificial intelligence is to build physical neural networks using emerging technology[11]. For this goal, spintronic nano-devices have key advantages, including their multifunctionality, fast dynamics, small size, low power consumption, high cyclability, high reliability and CMOS compatibility[12,13]. Furthermore, the high-speed dynamics of spintronic devices provides them key features for the emission, reception and processing of RF signals[14–20]. Several studies have shown their potential for building hardware neural networks[11,21–25]. In particular, it was recently proposed to use the flagship devices of spintronics, magnetic tunnel junctions (MTJs) as synapses taking RF signals as inputs[26–28], and neurons emitting RF signals at their output[23].

In this study, we first experimentally demonstrate that MTJs can perform synaptic weighted sums on RF signals containing multiple frequencies, similar to real-life RF data. Next, to showcase the potential of MTJs in RF signal classification, we construct a neural network using experimental data from MTJs acting as both synapses and neurons. We employ a backpropagation-free method know as extreme learning[29,30] to train this network, integrating both experimental results and software processing. We classify analogue RF signals encoding noisy four-pixel images, into three classes with 99.7 % accuracy, and into six classes with 93.2 % accuracy, which is as good as the equivalent software network. These results open the path to embedded systems performing artificial intelligence at low energy cost and high speed on complex RF signals, without digitization.

**Analogue processing of multiple radiofrequency signals in parallel**

We natively process analog RF signals by leveraging the intrinsic fast dynamics of nanodevices called magnetic tunnel junctions (MTJs). These devices are nanopillars composed of two ferromagnetic layers separated by a tunnel barrier. When an RF current is injected into an MTJ, as depicted in Figure 1, the magnetization of one layer enters in resonance with the input signal and by magnetoresistive effect a direct voltage is generated[31]. This phenomenon, called spin-diode, is frequency selective: the output voltage is only generated when the input signal is close to the resonance frequency of the device. We first illustrate this effect by sending single frequency RF signals at the input of our junctions. Figure 1 depicts how four MTJs of different resonance frequencies each process the spectrum from 100 MHz to 800 MHz. All four devices are from a material stack of $SiO_2$ // 5 Ta / 50 CuN / 5 Ta / 50 CuN / 5 Ta / 5 Ru / 6 IrMn / 2.0 $Co_{70}Fe_{30}$ / 0.7 Ru / 2.6 $Co_{40}Fe_{40}B_{20}$ / MgO / 2.0 $Co_{40}Fe_{40}B_{20}$ / 0.5 Ta / 7 NiFe / 10 Ta / 30 CuN / 7 Ru, where thicknesses are indicated in nm, and have different diameters from 250 nm to 450 nm. The typical resistance area product of the devices is 8 $\Omega\mu m^2$. Using individual magnetic fields, we can fine tune the frequencies of the devices.

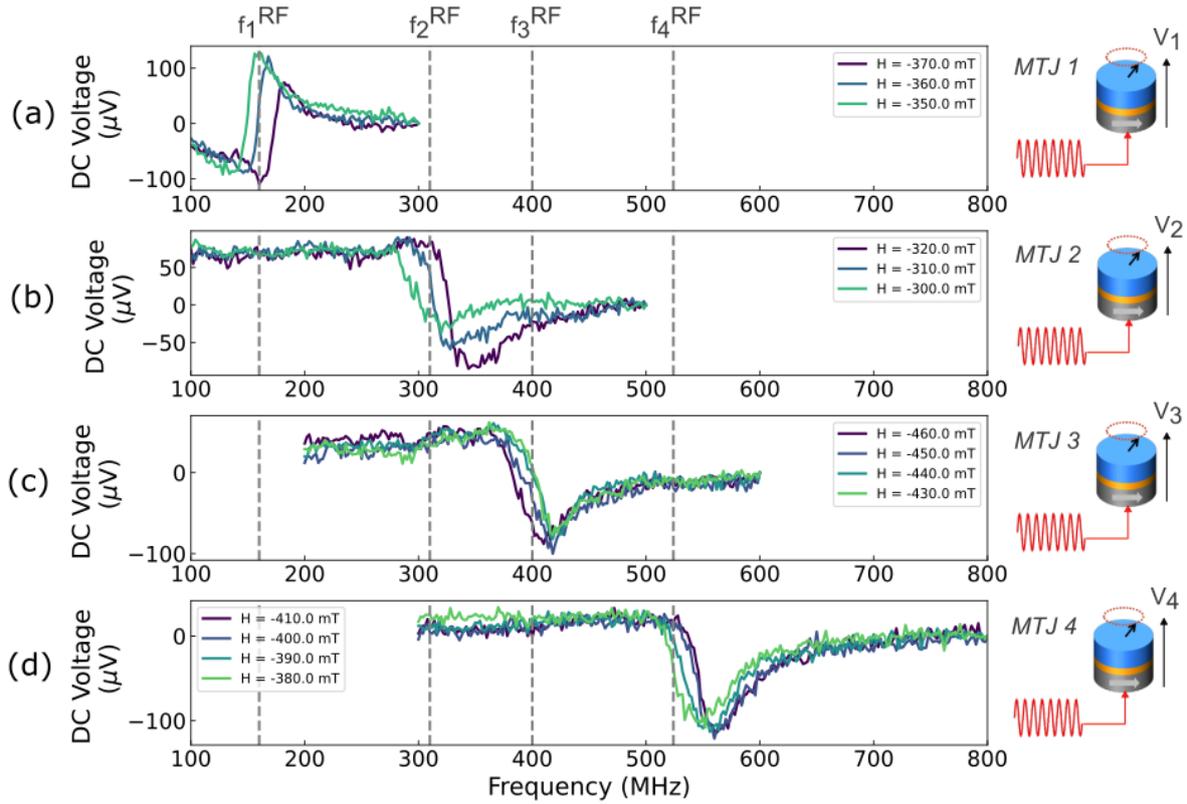

*Figure 1. (a-b-c-d) Rectified DC voltage versus input RF frequency for the four magnetic tunnel junctions. The input RF power is 10 µW. For each device, the measurement is repeated for different external magnetic fields. The field is applied perpendicular to the plane of the device. Inset: schematic of a magnetic tunnel junction. The magnetic fields are, from yellow to purple: (a) 370 mT, 360 mT and 350 mT, (b) 320 mT, 310 mT and 300 mT, (c) 460 mT, 450 mT, 440 mT and 430 mT, (d) 410 mT, 400 mT, 390 mT and 380 mT.*

The output DC voltage $V_i$ is proportional to the input RF power $P_i$ injected in the device and can be expressed as:

$$V_i = P_i \times w_i(f_i^{RF} - f_i^{res}),$$

where $f_i^{RF}$ and $f_i^{res}$ and the input and resonance frequencies respectively. Each MTJ can therefore be seen as performing synaptic operations through the multiplication of the analog input RF signals by a weight $w_i$ which is a function of $f_i^{RF} - f_i^{res}$. The individual synaptic weights $w_i$ of each junction can thus be tuned through the resonance frequencies of the devices. Figure 2 illustrates this synaptic operation for each of the four junctions (each column studies a different junction). In order to have a significant output DC voltage, for each MTJ we chose an input frequency close to its resonance: $f_i^{RF} = 160\ MHz$, $f_i^{RF} = 310\ MHz$, $f_i^{RF} = 400\ MHz$ and $f_i^{RF} = 524\ MHz$ for each junctions respectively. The top row of Figure 2 shows that the output DC voltage is proportional to the input RF power, where the proportionality factor – the synaptic weight – is controlled by the magnetic field via the resonance frequency, and can be set both to positive and negative value. The bottom row compares the experimental voltage to the expected voltage for a perfect weight multiplication. The normalized root-mean-square errors below 10 % (2.5 %, 6.3 %, 7.7 % and 5.6 % of the range for each MTJs respectively), demonstrate the ability of the junctions to perform synaptic operations on single signal inputs.

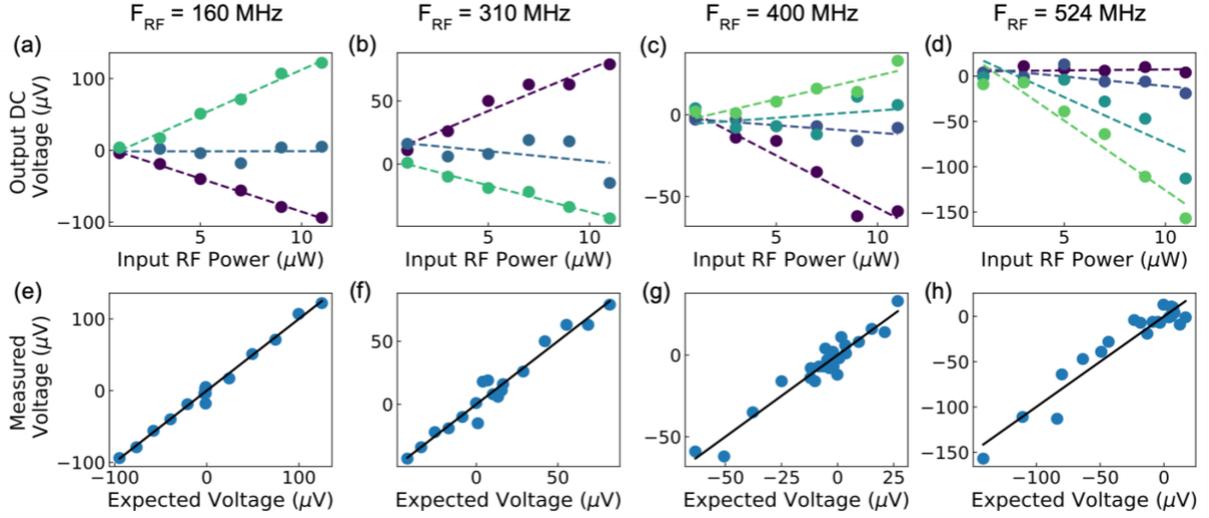

*Figure 2. (a-b-c-d) Synaptic multiplication: rectified DC voltage versus RF power, for different magnetic fields (colors). The dots are measurements while the dashed lines are linear fits. (e-f-g-h) Accuracy of the operation: measured voltage (dots) versus the ideal voltage (solid line). The magnetic fields are, green yellow to purple: (a) 370 mT, 360 mT and 350 mT, (b) 320 mT, 310 mT and 300 mT, (c) 460 mT, 450 mT, 440 mT and 430 mT, (d) 410 mT, 400 mT, 390 mT and 380 mT.*

As the RF signals interesting for applications are typically composed of several frequencies, we then study the response of the junctions when they each simultaneously receive the sum of the four RF inputs, as schematized in Figure 3(a). Analytically, the resulting output DC voltage of each MTJ can then be expressed as:

$$V_i = \sum_{k\ inputs} P_k \times w_{ik}(f_k^{RF} - f_i^{res}),$$

where $P_k$ and $f_k^{RF}$ are the power and frequency of each RF input signal. Figures 3(b-c-d-e) show the measured voltage versus the ideal expected voltage for each device, for all combinations of applied input powers and synaptic weights. The ideal voltages are computed using the individual synaptic weights from the responses to individual RF inputs. The fact that there is a good agreement (normalized root-mean-square errors of 8.0 %, 6.5 %, 8.3 % and 6.6 % respectively) between the measured and ideal voltages demonstrates the ability of the MTJs to linearly sum RF signals and function as synapses when they receive several RF inputs simultaneously.

In neural networks, the output of synapses connecting to a neuron are summed before a non-linear activation function is applied by the neuron[32]. In our case, the sum $V$ of the outputs of the four MTJs can be expressed as:

$$V = \sum_{\substack{i \\ MTJs}} V_i = \sum_{\substack{i \\ MTJs}} \sum_{\substack{k \\ inputs}} P_k \times w_{ik}(f_k^{RF} - f_i^{res})$$

$$V = \sum_{k\ inputs} P_k \times W_k$$

The total voltage $V$ is a weighted sum of the input powers by tunable weights, as desired. Now, each synaptic weight $W_k$ is encoded by all MTJs simultanesously, although the main contribution comes from the device whose resonance frequency is closest to the input frequency. Figure 3(f) shows that there is a good agreement (the normalized root-mean-square error is 4.4 % of the range) between the measured and expected summed voltage $V$ when the four sets of oinput powers and weights are varied. We observe that the agreement is better for the sum of all outputs than for the individual device outputs. This is because errors on the individual voltages are averaged out through the sum. As only the result of the sum is meaningful for the neural network, this is promising for the scalability of the system. These results demonstrate that arrays of MTJs can process multiple RF inputs simultaneously, over a wide frequency range in parallel. Although the sum of the voltage outputs from each junction has been performed here numerically for practicality, it can be achieved in a compact way on chip in the future by simply connecting them electrically[26].

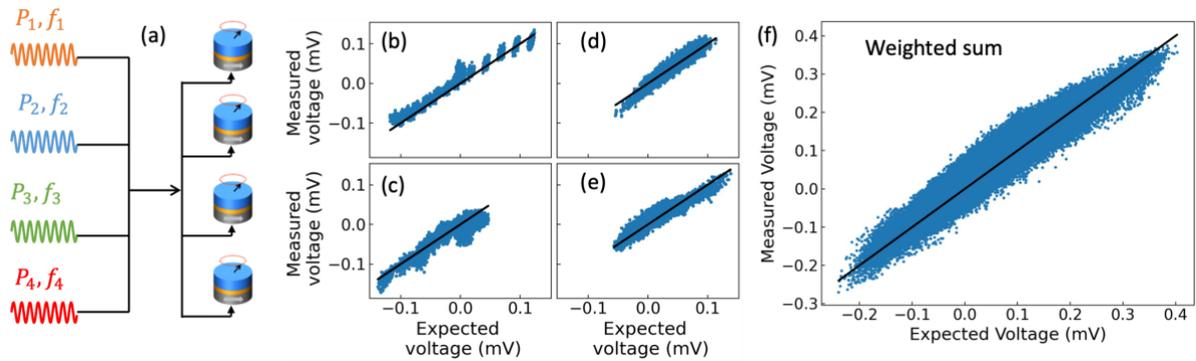

Figure 3. (a) Schematic of each MTJ simultaneously receiving the four RF inputs. (b-c-d-e) Measured voltage (dots) versus expected voltage (solid line), for each MTJ. For each plot, all combinations of input powers (from 1 to 11 µW by 2 µW steps) and weights values where measured. (f) Measured voltage (dots) versus expected voltage (solid line) for the sum. All combinations of the four sets of powers and weights were measured.

**Experimental classification of RF signals**

We now use the RF synaptic weighted sums to perform a classification task. Here we choose a method, sometimes called "extreme learning"[29,30], depicted in Figure 4. The neural network is composed of two fully connected layers, separated by a hidden layer of neurons. The first layer of synapses, described by a vector $W^{(1)}$ – here implemented in hardware – has random weights and is not tuned during training. The second layer of synapses, described by a vector $W^{(2)}$ – here implemented in software – is trained through a simple matrix inversion, as detailed below. Extreme learning is similar to reservoir computing[33] in the sense that there are random hardware weights which are not trained and a software layer of weights that are trained. However, in contrast to the reservoir approach, in extreme learning the network is feedforward and static, meaning there are no recurrences or connections between the neurons of the hidden layer. This method has the advantage of eliminating the need for backpropagation, enabling classification without adjusting the weights. Although extreme learning may not be suitable for tackling complex, state-of-the-art tasks, it is a good benchmark for artificial neural networks implemented with emerging hardware.

The equation describing the extreme learning neural network considered here is:

$$y = W^{(2)} a(W^{(1)} \times P),$$

where $y$ is the output vector, $a$ is the activation function of the neurons and $P$ is the vector of input powers. As usual in extreme learning[29,30], the weights of the second layer are chosen as:

$$W^{(2)} = \mathcal{F}\left(a(W^{(1)} \times P_{all})\right) \hat{y}_{all},$$

where $P_{all}$ is the matrix of all input vectors in the dataset, $\hat{y}_{all}$ is the matrix of all target outputs in the dataset and $\mathcal{F}$ is the pseudo-inverse function. The class of the signal is the output with the highest value.

We compose the first fully connected layer using the experimental weighted sums. By using all combinations of measured weights of the four MTJs (i.e. all combinations of the weights shown in Figure 2), we obtain 144 pseudo-random sets of weights. The results of the 144 MAC operations are injected into MTJ neurons. These MTJs are used as neurons as in [23], where the non-linear relationship between their output RF power and input DC current serves as activation function. We implement 144 activation functions with two MTJ devices. The two MTJs are measured under different conditions so to vary their activation function and thus mimic device to device variability: using DC current injection in a strip line above the device we generate a different local Oersted field for each neuron.

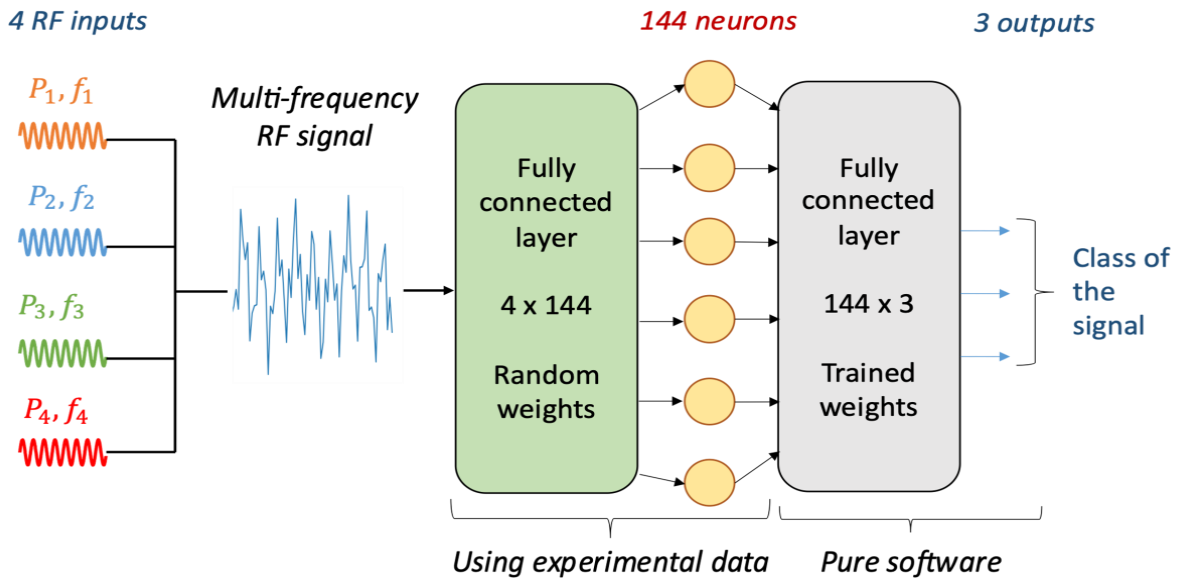

*Figure 4. Schematic of classification of multi-frequency RF signal using extreme learning.*

While the first layer of weights and the neurons are implemented with the experimental data, the second layer of weights and the matrix inversion are implemented in software. To benchmark our experimental network, we also perform classification with an equivalent software network, where the first layer is composed of ideal weighted sums with weights extracted from the experiment, and the neurons are conventional rectified linear units.

We compose a dataset of analogue RF signals as follows. Each sample of the dataset is a four-pixel image, as the ones shown in Figure 5(a). Each pixel corresponds to a frequency ($f_i^{RF} = 160\ MHz$, $f_i^{RF} = 310\ MHz$, $f_i^{RF} = 400\ MHz$ and $f_i^{RF} = 524\ MHz$) and the intensity of the pixel is encoded by the RF power at that frequency. In order to emulate noise in the inputs, we assign the powers 1 and 3 μW to the gray pixels and 7 and 9 μW to the black pixels.

We first evaluate the ability of the network to discriminate the three classes shown in Figure 5(a). We perform 100 runs, with 4 randomly chosen samples per class for the training set, and 20 randomly chosen samples per class for the test set. We obtain 99.7 % for the experimental network and 96.2 % for the equivalent software network, with standard deviations of 0.9 % and 7.2 % respectively. If we complexify the task by having six classes (all possible combinations of two chosen pixels in the image), the test accuracy becomes 93.2 % for the experimental network and 94.9 % for the equivalent software network, with standard deviations of 3.6 % and 5.2 % respectively. Figure 5(b-c-d-e) shows the corresponding confusion matrices for both tasks. These results demonstrate that a network composed of experimental RF MTJs data can classify raw analogue RF signals, with accuracy as high as a software network.

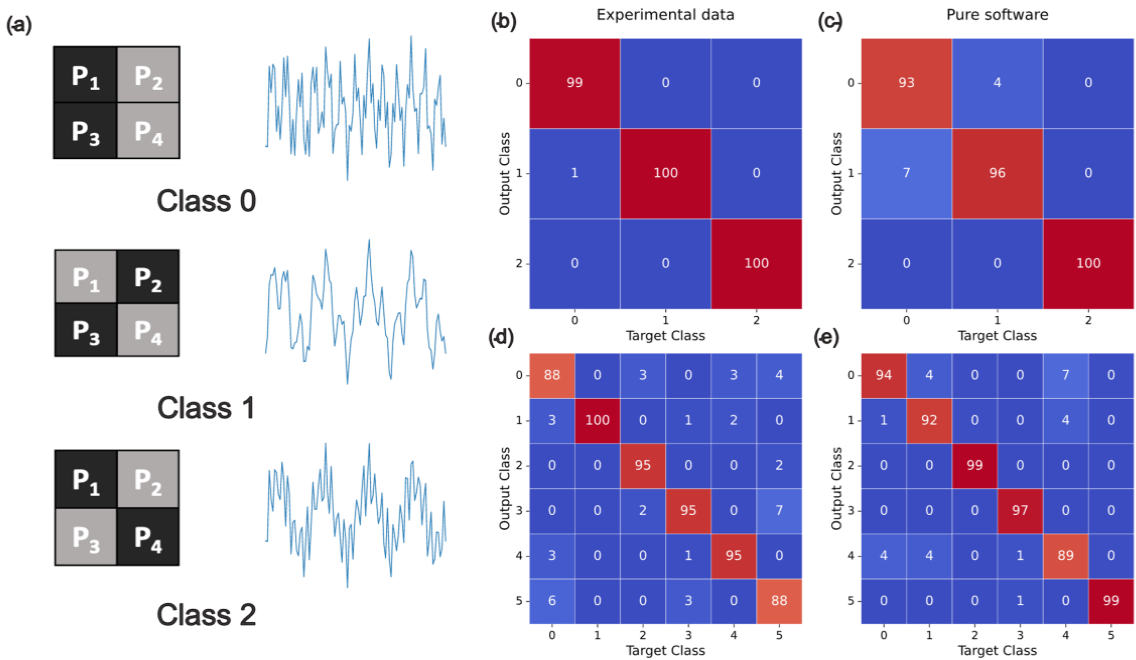

Figure 5. (a) The three classes of signals as four-pixel images, each accompanied with a sample represented as a time varying signal. (b-c-d-e) Confusion matrices for the classification of RF signals, both for the purely software network (c-e) and for the network using experimental data (b-d), for the three-class (b-d) and six-class (c-e) tasks. The labels indicate the percentage of samples from each target class that is classified into each output class.

## Conclusions

We have leveraged the dynamics of magnetic tunnel junctions to perform synaptic operations, and performed weighted sums on several analogue RF signals in parallel. In the future, by choosing the materials, shape and size of the devices, their frequency can be engineered from 50 MHz to 50 GHz[34]. As a consequence, an array of magnetic tunnel junctions could process RF signals over this whole frequency range in parallel, without digitization. This removes the need of multiple local oscillators or high-speed

ADCs[35]. Using experimental data from RF MTJs functioning as both synapses and neurons, we have composed a neural network and demonstrated classification of RF signals through extreme learning. The achieved accuracy is on par with that of an equivalent software-based network. These results open the path towards large neural networks able to perform artificial intelligence tasks on raw RF signals, without digitization, at low energy cost and small size.

**Acknowledgments**

This work has received funding from the European Union under grant PADR – 886555-2 – SPINAR.